**ORIGINAL ARTICLE**

# Milton Friedman's spending matrix revisited: 'Spending efficiency' and 'preference compatibility' across different economic systems

## Ali Zeytoon-Nejad

Wake Forest University, Winston-Salem, North Carolina, USA

**Correspondence**
Email: zeytoosa@wfu.edu

**Abstract**

This article expands Milton Friedman's spending matrix to analyse 'spending efficiency' and 'preference compatibility' across different economic systems against five key outcome criteria. By generalising Friedman's typology, it compares efficiency and freedom as systems shift from laissez-faire capitalism to communism, illustrating a gradual deterioration in their key outcomes. While government intervention is sometimes necessary to address market failures, its role should always be carefully limited to avoid inefficiency and misalignment with individual preferences. The insights may provide guidance for policymakers in designing economic systems and policies that promote both economic prosperity and personal liberty.

# 1 | INTRODUCTION

The concepts of 'economic efficiency' and 'preference compatibility' have long attracted the attention of economists and policymakers. Scholars such as Amartya Sen, in his seminal work *Development as Freedom* (1999), emphasise the critical importance of aligning economic policies with individual freedoms and preferences to achieve true development. Joseph Stiglitz, in *Economics of the Public Sector* (1986), discusses how government interventions often fail to achieve the desired efficiency due to misaligned incentives and bureaucratic inefficiencies. Central to this discourse is the framework proposed by Milton Friedman, whose spending matrix offers a clear illustration of how different spending mechanisms can impact economic outcomes. However, Friedman's original model, while insightful, operates under certain implicit assumptions that limit its applicability to broader real-world scenarios. This article seeks to extend Friedman's spending model by relaxing these assumptions to encompass a broader array of spending mechanisms observed across various economic systems.

In the realm of economic theory, the transition from laissez-faire capitalism to pure communism represents a spectrum along which various spending mechanisms can be analysed. The extension introduced in this article allows for a logical comparison of the efficiency and preference compatibility of different economic systems and various spending mechanisms, including the laissez-faire economic system, controlled economies with regulations, welfare-state economies, the communist system, as well as spending mechanisms such as paternalistic spending, systems of income transfers, donations or financial aid with and without conditions, and coupon- and quota-based distribution mechanisms. The article evaluates each of these economic systems and spending scenarios against five key outcome criteria: preference compatibility for the giver, efficient use of resources, removal of intermediaries, market efficiency, and preference compatibility for the taker.

My analysis posits that as an economic system shifts along this spectrum, there is a notable and systematic decline in both spending efficiency and preference compatibility. The analysis reveals an alarming trend: as economic systems move away from the principles of laissez-faire capitalism, not only does spending efficiency deteriorate but the alignment of spending with individual preferences also weakens. This observation has profound implications for policymakers striving to design economic systems and policies that aim to promote economic performance and societal welfare, emphasising the necessity of preserving economic freedom to ensure both efficiency and satisfaction.

In synthesising aspects of the ideas of two economics Nobel Laureates, Milton Friedman and Friedrich von Hayek, this article underscores a critical insight, namely that the erosion of spending efficiency and preference compatibility not only hampers economic performance but also threatens the fundamental freedoms essential to a prosperous society. By emphasising the importance of maintaining both efficiency and freedom, the study contributes to the ongoing debate on how to best design economic systems that foster sustainable and just growth. This article aligns with Hayek's view by exploring how economic systems that maintain higher levels of individual freedom and less government intervention, such as free-market capitalism, exhibit superior spending efficiency and preference compatibility. Conversely, systems with more centralised control and less economic freedom tend to suffer from inefficiencies and misalignments with individual preferences.

The following sections of the article are organised as follows: Section 2 provides a brief literature review, examining foundational and recent contributions to economic thought that relate to spending efficiency and preference compatibility. Section 3 delves into the core framework of Friedman's spending matrix, explaining the four types of spending and their implications for economic efficiency. Additionally, five criteria are introduced to delve deeper into the characteristics of each type of spending and their logical economic outcomes. Section 4 expands the original spending-typology model to include additional spending mechanisms and thereby broadens the analysis. This section also presents the main discussion and provides final thoughts, integrating the extended model with real-world applications and theoretical insights, and providing an examination of how various spending mechanisms perform across different economic systems. Finally, section 5 concludes and summarises the article. It also synthesises the key findings and highlights the practical implications for policymakers and economists.

## 2 | MAJOR CONTRIBUTIONS TO THE LITERATURE

Milton Friedman's contributions to economic thought are profound, spanning various domains from consumer theory and monetary theory to public policy and comparative economic systems. One of his seminal works, *Capitalism and Freedom* (1962), articulates the core principles of economic freedom and its interdependence with political freedom. Friedman argues that economic freedom is not only an end in itself but also a necessary means to political freedom, as it limits the concentration of power and promotes individual autonomy. Friedman's analysis of government spending is particularly relevant to this study. In his critique of governmental intervention, Friedman (1962) contends that public spending often lacks the efficiency of private spending due to the separation of spending and benefit recipients, which leads to inefficiencies and misaligned incentives. This article takes this critique further by examining not only governmental spending but also other forms of spending, such as paternalistic spending, income transfers, donations or financial aids with and without conditions, and coupon-based and quota-based allocation mechanisms. The present article builds on Friedman's assertion by examining how different economic systems, through their spending mechanisms, impact both economic efficiency and individual freedom.

In another influential work, *Free to Choose* (1980), co-authored with his wife Rose, Friedman expands on the concept of personal choice as a driving force behind economic efficiency. The book emphasises that when individuals spend their own money on themselves, they are likely to be most efficient, as they have both the incentive to economise and the knowledge of their own preferences. This principle is a cornerstone of Friedman's spending matrix, which categorises spending into four types based on who is spending the money and on whom it is being spent. The current article relies on this typology to compare and contrast spending efficiency and preference compatibility within Friedman's four spending scenarios, and also extends his typology to incorporate additional spending mechanisms and explores how the efficiency and preference compatibility of these mechanisms vary across different economic systems.

Robert Nozick's *Anarchy, State, and Utopia* (1974) emphasises individual agency and critiques government's income redistribution as a violation of property rights that distorts voluntary market exchanges and reduces economic autonomy. Buchanan and Tullock's *The Calculus of Consent* (1962) and Downs's *An Economic Theory of Democracy* (1957) further complement this analysis by showing how government spending decisions are often shaped by self-interested political incentives, resulting in inefficiencies and misaligned outcomes. Together, these works provide a rich theoretical foundation for the current article's argument that economic systems function more effectively when spending decisions are decentralised and closely aligned with individual preferences.

Friedrich von Hayek's contributions to economic theory, particularly regarding the role of knowledge and the dangers of central planning, provide critical context for understanding the implications of spending efficiency and preference compatibility in different economic systems. In *The Road to Serfdom* (1944), Hayek argues that central planning leads to the erosion of individual freedoms and the creation of economic inefficiencies resulting from the inability of a central authority to possess and use all necessary information for effective decision-making. Hayek's exploration of spontaneous order and the price mechanism in his article 'The Use of Knowledge in Society' (1945) underscores the importance of dispersed knowledge in achieving economic efficiency. He posits that the price system enables individuals to use localised knowledge, thereby coordinating their actions more effectively than any central planner could do through a process of aggregate decision-making. In *Rules and Order* (1973), Hayek further elaborates on the importance of a legal framework that supports individual freedom and limits government intervention. He contends that overreaching legislation can stifle innovation and economic growth.

These principles are closely related to those of Friedman. This article builds upon Hayek's critique by examining how centralised spending mechanisms, such as governmental and paternalistic spending, perform weakly in terms of efficiency and preference compatibility compared with decentralised mechanisms like the laissez-faire capitalist economic system.

The examination of spending efficiency and preference compatibility has garnered significant attention from scholars in the fields of development economics and public economics as well. One such influential work is Amartya Sen's *Development as Freedom* (1999), where Sen argues that development should be assessed by the expansion of freedoms rather than merely by economic growth indicators. Sen emphasises the role of individual agency and the importance of creating conditions where individuals can freely pursue their goals. Joseph Stiglitz's *Economics of the Public Sector* (1986) provides a comprehensive analysis of government intervention in the economy, including the role of public spending. Stiglitz argues that while government spending can address market failures, it often suffers from inefficiencies resulting from bureaucratic constraints and lack of proper incentives.

Recent institutional scholarship has continued to explore the themes of spending efficiency and preference compatibility, providing new insights and empirical evidence. Daron Acemoglu and James Robinson's *Why Nations Fail* (2012) analyses how inclusive economic and political institutions, which consider individual preferences and freedoms, foster economic prosperity and efficiency. Acemoglu and Robinson argue that the alignment of institutions with individual preferences is crucial for achieving sustainable economic growth.

By extending Friedman's spending model and incorporating insights from these classic studies, this article aims to fill in a gap in the prior literature by providing a nuanced framework for comparing the efficiency and preference compatibility of different economic systems. This approach not only enriches the theoretical foundation but also offers practical implications for policymakers seeking to optimise economic performance and social welfare. I aim to provide a nuanced understanding of why market-oriented systems tend to maintain both efficiency and individual freedom while fostering economic growth, which also contributes to the broader understanding of institutional effectiveness.

## 3 | FRIEDMAN'S QUADRANTS: FOUR WAYS OF SPENDING

This section delves into the theoretical underpinnings of Milton Friedman's spending model and explains them in relation to five key outcome criteria. Friedman categorises ways of spending into four distinct types based on 'who is spending the money?' and 'who benefits from the spending?'. This framework is instrumental in understanding the varying degrees of spending efficiency and preference compatibility under different economic systems. Friedman's typology of spending can be structured in a matrix comprising four distinct quadrants. These quadrants are as follows:

Quadrant 1: *Individuals spending their own money on themselves*. In this quadrant, individuals spend their own money to purchase goods or services for themselves. For Friedman, this type of spending is the most efficient and most preference-compatible. People have a strong incentive to economise and maximise utility because they are directly aware of their preferences and the cost of the goods or services they are purchasing. The close alignment of expenditure and benefit ensures minimal waste and optimal satisfaction.

Quadrant 2: *Individuals spending their own money on others*. Here, individuals use their money voluntarily to buy goods or services for others. While there is still an incentive to economise as they are spending their own funds on others, the preference compatibility is somewhat reduced compared with the first quadrant. This is because the spender may not fully understand the preferences of the recipient, potentially leading to a mismatch between the spending and the recipient's actual needs or desires. Although still relatively efficient, this type of spending introduces some preference incompatibility due to the lack of complete information about the recipient's true preferences, which may arise due to imperfect communications, lack of a mutual understanding, or the nature of making one single aggregate decision for a whole, among others.

Quadrant 3: *Individuals spending other people's money on themselves*. In this quadrant, individuals spend money that is not their own to benefit themselves. This situation often arises in organisational or bureaucratic settings. The primary inefficiency here stems from the reduced incentive to economise, as the spender does not bear and may not appreciate the cost of the expenditure. While the spender has a good understanding of their own preferences, the lack of personal financial responsibility can lead to overspending or less cost-effective choices, reducing overall spending efficiency.

Quadrant 4: *Individuals spending other people's money on others*. This is, according to Friedman, the least efficient and the least preference-compatible type of spending. It typically occurs in governmental contexts where decision-makers allocate resources that belong to others (taxpayers) for the benefit of third parties. The dual disconnect – both the original contributor and the final beneficiary of the funds being different entities from the spender of the funds – leads to significant inefficiencies in spending, usually through a compounded principal–agent problem involving multiple layers. The spender has less incentive to economise and usually lacks comprehensive knowledge of the beneficiaries' true preferences or makes one single aggregate decision for a whole which does not necessarily match all individual preferences, often resulting in suboptimal allocation of resources as well as a huge loss of efficiency due to preference mismatches.

Friedman's spending typology can be summarised in a matrix form in terms of 'you' and 'someone else' in the way depicted in Figure 1.

Traditionally, Friedman's four ways of spending have been discussed primarily in terms of 'cost economisation' and 'value maximisation'. However, there is significantly more depth and breadth to this model that extends well beyond these traditional parameters. By examining the model through the lenses of various concepts of efficiency and different notions of preference compatibility, we can uncover a richer understanding of the economic dynamics at play. This broader perspective would reveal the multifaceted implications of spending mechanisms and offer insights into how different types of spending can influence overall economic efficiency and alignment with individual preferences.

## *Ways of Spending Money*

### Whose money is spent?

|  |  | Yours | Someone else's |
|---|---|---|---|
| **On whom is money spent?** | **You** | **Quadrant 1**<br><br>You spending your own money on yourself | **Quadrant 3**<br><br>You spending someone else's money on yourself |
|  | **Someone else** | **Quadrant 2**<br><br>You spending your own money on someone else | **Quadrant 4**<br><br>You spending someone else's money on someone else |

**FIGURE 1** Milton Friedman's spending typology.

We can examine the characteristics of each quadrant in terms of five outcome criteria:

1. *Contributor's preferences: Preference-compatibility for the earner/giver/contributor.* This criterion evaluates how well the spending aligns with the preferences and possibly moral values of the earner, contributor, or donor, who is sometimes called 'the giver' here for simplicity. When individuals spend their own money, they can direct it towards causes and activities that resonate with their personal priorities and values. In contrast, when entities such as the government decide on spending priorities on behalf of an individual, the allocation may not align with the giver's personal and moral preferences, leading to potential dissatisfaction.

2. *Utilisation efficiency: Efficient use of the transferred resource.* This criterion focuses on the efficient use of resources, emphasising 'minimal waste' and 'optimal usage'. Individuals who earn, spend and consume their own money are more likely to use resources efficiently because they have a direct stake in the outcome. This personal investment encourages careful spending and reduces the likelihood of waste, which ensures that resources are used in the most efficient and effective manner.

3. *Intermediation cost efficiency: Efficiency gain from removing the intermediary.* This criterion addresses the efficiency gained by eliminating intermediaries, thereby reducing the costs associated with intermediation, bureaucracy, and potential corruption. Direct transactions between private parties, whether in a market setting or through bilateral negotiations, tend to be more transparent, more cost-effective, and more efficient, avoiding the leakages,[1] delays and potential corruption often associated with bureaucratic processes.[2] Removing intermediaries minimises such inefficiencies and can reduce the risk of fraudulent practices.

4. *Market efficiency: Efficiency in minimising market distortions and potential deadweight losses.* This criterion assesses the impact of spending on market efficiency and the total economic surplus in society. Market distortions, such as those caused by taxes and subsidies, can lead to reduced trade opportunities, reduced gains from trade, and deadweight losses, as discussed in welfare economics.[3] When spending decisions are made by private entities in a decentralised manner, market forces can more effectively regulate transactions, leading to optimal bargaining outcomes, free-market equilibria, and minimal or no deadweight losses. This ensures a higher level of market efficiency and preserves the total surplus, a measure of society's well-being.

5. *Recipient's preferences: Preference-compatibility for the recipient/taker/beneficiary.*[4] This criterion evaluates how well the spending meets the actual needs, values, and preferences of the recipient, who is sometimes called 'the taker' here for simplicity. When individuals receive funds directly, they have the autonomy to spend them according to their specific needs and priorities. In contrast, government-directed spending may impose restrictions or guidelines that do not align with the recipient's preferences, potentially leading to suboptimal satisfaction and reduced utility due to a lack of freedom of choice.[5]

These five criteria provide a comprehensive framework for analysing the four quadrants of Friedman's spending typology matrix, which can in turn have implications for analysing the characteristics and outcomes of different spending mechanisms within various economic systems.

We can now examine the likely outcome of each way of spending (quadrant) in terms of the five key criteria introduced above.

*Outcome of Quadrant 1: Individuals spending their own money on themselves.* In this quadrant, individuals spend their own money to purchase goods or services for themselves. This type of spending is highly preference-compatible for the giver, as individuals can direct their resources towards what they value most according to their personal and moral preferences. The efficient use of the transferred resource is maximised because individuals have a strong incentive to economise and ensure minimal waste; their direct stake in the expenditure and consumption encourages them to make the best use of their money. Additionally, this type of spending eliminates the need for intermediaries, avoiding the costs and inefficiencies associated with intermediation, such as administrative expenses and potential corruption. Market distortion and deadweight loss are non-existent or minimised since the transactions occur directly between buyers and sellers without external interference, preserving market efficiency and total surplus. Lastly, this quadrant ensures high preference-compatibility for the taker, as individuals spend their own resources in ways that best meet their own needs and priorities, leading to maximum satisfaction and utility.

*Outcome of Quadrant 2: Individuals spending their own money on others.* Here, individuals spend their own money voluntarily to purchase goods or services for others. While this type of spending remains efficient overall, preference compatibility for the recipient may be reduced compared to Quadrant 1 because the giver, who is spending the money, may not fully understand the recipient's true preferences, which can lead to a preference mismatch in spending. However, the giver still has a personal incentive to economise and avoid waste, ensuring that the resources are used efficiently. The removal of intermediaries continues to provide efficiency benefits by reducing administrative costs and potential corruption. Market distortion and deadweight loss are still minimised as private transactions predominate, maintaining market efficiency and maximising total surplus and social welfare. Overall, in this quadrant preference compatibility for the taker may be reduced, as recipients may not always receive what they most need or desire, but this method of spending can still offer significant benefits when the giver has some understanding of the recipient's preferences and agrees to act accordingly.

*Outcome of Quadrant 3: Individuals spending other people's money on themselves.* In this case, individuals spend money that is not their own for their own benefit. Preference compatibility for the initial giver is very low, as they can no longer choose on which goods and services the money must be spent such that the spending aligns with their own preferences and priorities. Additionally, the efficient utilisation of the transferred resource is compromised because the recipient, who is spending the money, does not bear or may not appreciate the cost, leading to less incentive to economise and potentially resulting in greater waste. As there is no intermediary in this spending scenario, there is no loss of efficiency due to additional administrative costs and the potential for corruption. Market distortion and deadweight loss cannot occur since the spending is not influenced by subsidies or other external factors which can reduce overall market efficiency and total surplus. Additionally, preference compatibility for the taker is still inherently high since the taker is also the spender and can easily align their spending with their own preferences.

*Outcome of Quadrant 4: Individuals spending other people's money on others.* This quadrant involves individuals spending money that is not their own for the benefit of others, often seen in governmental contexts or charitable organisations. Preference compatibility for the giver is low, as the intermediary (e.g. the government) may make allocation decisions that do not reflect the giver's own personal preferences or moral values. The efficient use of resources is significantly challenged due to the intermediary's lack of personal financial responsibility and the absence of a vested interest in the resource allocation and lack of stake in the final consumption. Besides, intermediaries play a significant role, adding layers of administrative costs and bureaucratic waste, as well as increasing the risk of corruption, altogether further reducing efficiency. Market distortion and deadweight loss are likely to be high because government interventions and subsidies and taxes can deter trade, disrupt market dynamics, reduce economic activities and transactions, and thereby lead to market inefficiencies and a reduced total surplus. Preference compatibility for the taker is often low as well, as recipients may be subjected to spending decisions that do not align with their actual needs or desires, which results in reduced satisfaction and utility.

To summarise all this information, we can introduce a coding system. As shown in Figure 2, this system uses five positive/negative signs within parentheses, corresponding to the five criteria discussed above. A positive sign indicates a desirable outcome concerning the corresponding criterion, while a negative sign indicates an adverse outcome with respect to that criterion.

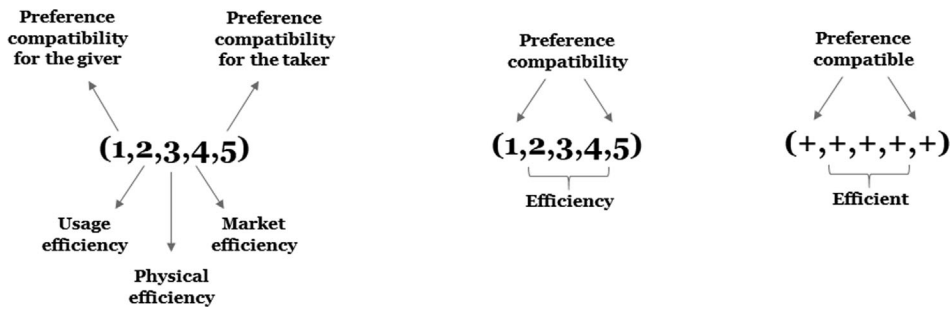

**FIGURE 2** A system of coding information about 'spending efficiency' and 'preference compatibility' for Friedman's spending typology.

Accordingly, all this information about spending quadrants and their respective characteristics in relation to the five discussed outcome criteria can now be summarised and organised in the matrix form shown in Figure 3.

As demonstrated in Figure 3, Quadrant 1, where individuals spend their own money on themselves, is both highly efficient and fully preference-compatible because individuals have a strong incentive to maximise utility and minimise waste, as they directly benefit from their spending decisions. This alignment of incentives ensures that resources are used effectively to achieve maximum efficiency, freedom, and utility, making it the ideal quadrant to operate in whenever possible. Quadrant 2, where individuals spend their own money on others, remains efficient but is somewhat preference-incompatible. While the giver is the one spending the money and may have different preferences from the recipient's preferences, the recipient's needs may not be fully met, leading to potential preference mismatches, though this type of spending is still relatively good overall if one takes into account all the five outcome criteria. In Quadrant 3, where individuals spend other people's money on themselves, efficiency drops as there is less incentive to economise, and it becomes somewhat preference-incompatible due to potential misalignment between the spender's desires and the original owner's intentions, making it not so good and less desirable as a way of spending. Quadrant 4, where individuals spend other people's money on others, is the least efficient and most preference-incompatible. This situation often involves bureaucratic processes that disconnect the spender from both the source of the funds and the recipients' needs, leading to significant inefficiencies and preference misalignments, making it the worst quadrant and the least desirable way of spending.[6]

In sum, Friedman's four quadrants of spending provide a comprehensive framework for understanding the varying degrees of efficiency and preference compatibility under different types of spending scenarios. Each quadrant highlights the unique challenges and advantages associated with who is spending the money and who is benefiting from it. By examining the five

## Ways of Spending Money

**FIGURE 3**  Friedman's spending typology in terms of its four quadrants and spending efficiency and preference compatibility of each quadrant.

outcome criteria – preference compatibility for the giver, efficient use of resources, removal of intermediaries, market efficiency, and preference compatibility for the taker – we can gain deeper insights into the economic dynamics at play. However, Friedman's typology of ways of spending contains several implicit assumptions that, if generalised, can reveal a more nuanced spectrum of spending mechanisms across different economic systems. By extending this typology, it is possible to illustrate how the characteristics of different spending methods evolve along a continuum, reflecting the diverse economic contexts in which they operate. This broader approach allows for an in-depth analysis of how key factors – such as efficiency, preference compatibility, incentive structures, information asymmetry, and administrative costs – vary within each economic system. This generalisation can enrich our understanding of the dynamic interplay between spending practices and their possible economic outcomes, which also highlights the usefulness and applicability of Friedman's original framework.

# 4 | AN EXTENSION OF FRIEDMAN'S SPENDING TYPOLOGY

While Friedman's spending typology provides an important framework for analysing spending efficiency and preference compatibility under different spending contexts, real-world economic systems introduce additional aspects and complexities. This section extends Friedman's model to include several additional spending mechanisms such as paternalistic spending, systems of income transfers, donations or financial aid with and without conditions, and coupon- and quota-based distribution mechanisms into this model. The extension also makes adjustments to the original model such that the resulting model can be used to make logical inferences about the relationship between the morphology of economic systems and their likely outcomes with respect to efficiency and preference compatibility. Thereby, the extended version will broaden the scope of analysis and provide deeper insights into the questions at hand. By doing so, we can assess how these different spending mechanisms perform in terms of efficiency and preference compatibility across diverse economic contexts and systems. This extended framework will also explore implicit assumptions behind the original model, explicitly address them in the extended version, present a generalised model that accounts for a wider range of real-world scenarios, and apply it to analyse spending mechanisms under a wide range of economic systems. Ultimately, this nuanced approach aims to provide valuable insights for economists and policymakers seeking to optimise spending efficiency and preference compatibility in various economic environments.

Friedman's spending typology is a two-dimensional model that tries to consider the roles of three parties in relation to earning (owning), spending, and consuming. While concise, the model's inability to explicitly address the interactions among these three parties is a significant shortcoming. For instance, in the first quadrant, a single individual performs all three roles. In the second and third quadrants, two parties are involved. In the fourth quadrant, three separate parties are involved: one person owns the money, a second person benefits from the spending, and a third party (such as the government) decides how the money should be spent. Additionally, in Quadrant 2 the giver is the spender, while in Quadrant 3 the taker is the spender, while all these are assumed implicitly in the model. Furthermore, the original model provides little or no information about whether the recipient has a say in choosing on what items the money is to be spent. To accommodate a wider range of scenarios involving earners, spenders, and consumers and all other aspects addressed above, an extended version of Friedman's typology is necessary. This extended model will enable us to remove such potential ambiguities and allow us to more accurately compare economic systems and spending mechanisms when evaluating them based on the three efficiency concepts and the two preference-compatibility notions introduced in this article. This section is to undertake this important task.

In the sign determinations reported in Figure 3, implicit assumptions are made, particularly for Quadrants 2 and 3. For instance, in Quadrant 2, it is assumed that the giver decides how the money is spent, which affects the preference compatibility for the recipient negatively, as indicated by the last sign in the parentheses. However, in reality, the recipient may sometimes have the autonomy to decide how to use the transferred money. For example, scholarships provided by private entities often allow recipients to choose their field of study, which would turn the last sign in the parentheses positive. Conversely, if the donor restricts the use of funds to a specific field of study, preference compatibility for the recipient remains negative. Similarly, in Quadrant 3 it is assumed that someone's money is spent directly by the recipient, while it could also be spent indirectly through an intermediary such as a government. Both factors – having or not having a choice in spending, and the presence or absence of an intermediary – significantly influence the five criteria of efficiency and preference compatibility. It is therefore important to consider these determinants when assessing the efficiency and preference compatibility of spending mechanisms in an economy. Figure 4 generalises this classification and examines the status of the five criteria under 12 possible scenarios.

Figure 4 presents an extended version of Friedman's spending typology by considering two additional attributes: the mechanism of transfer (direct or indirect) and discretion in consumption (presence or lack of freedom of choice). The mechanism of transfer refers to whether the funds are allocated directly by the spender or through an intermediary, such as a government or organisation. Discretion in consumption indicates whether the recipient has the freedom to choose how to spend the funds or if there are specific restrictions imposed by the giver. Accordingly, 12 cases can be defined, as shown in Figure 4. Below, some of these cases are further explained and each will be described as analogous to a particular spending scenario or an economic system.

## 4.1 | Case 1: Direct spending with full discretion

### 4.1.1 | Description

In this scenario, you spend your own money and have complete freedom to choose how to allocate it. This means you have full control over the spending decisions, ensuring that the expenditures align perfectly with your preferences and priorities. This type of spending exemplifies the highest degree of personal autonomy and accountability.

### 4.1.2 | Criteria evaluation

1. *Preference compatibility for the giver.* As the earner and spender, you are willingly spending your money and can also direct your resources towards what you value most, ensuring a high level of satisfaction and alignment with your personal preferences and moral values. (Positive.)

## Ways of Spending Money

### Whose money is spent?

| | | Yours | Someone else's | |
| --- | --- | --- | --- | --- |
| | | | Directly | Indirectly (Intermediary, e.g. gov.) |
| You | Can choose on what to spend | **Case 1:** Freedom of Choice (Efficient and Preference-Compatible) **(+,+,+,+,+)** | **Case 5:** Lack of stake in usage **(+,-,+,+,+)** | **Case 9:** Preference-compatible for the recipient but not for the giver **(-,-,-,-,+)** |
| | Cannot choose on what to spend | **Case 2:** Lack of freedom of choice in usage **(+,+,+,+,-)** | **Case 6:** Lack of freedom of choice and stake in usage **(+,-,+,+,-)** | **Case 10:** Coercion of choice and preference-incompatibility for all **(-,-,-,-,-)** |
| Someone else | Can choose on what to spend | **Case 3:** Lack of stake in usage **(+,-,+,+,+)** | **Case 7:** Lack of stake in usage **(+,-,+,+,+)** | **Case 11:** Preference-compatible for the recipient but not for the giver **(-,-,-,-,+)** |
| | Cannot choose on what to spend | **Case 4:** Lack of freedom of choice and stake in usage **(+,-,+,+,-)** | **Case 8:** Lack of freedom of choice and stake in usage **(+,-,+,+,-)** | **Case 12:** Coercion of Choice (Inefficient and Preference-Incompatible) **(-,-,-,-,-)** |

*On whom is money spent?*

**FIGURE 4** An extended version of Friedman's spending typology in terms of spending efficiency and preference compatibility under 12 scenarios.

2. *Efficient use of the transferred resource.* Given that you are the earner, the spender, and the end consumer in this scenario, then you have a full appreciation of the effort required to earn the money and also have a direct financial stake in the outcome. Therefore, you have a strong incentive to use the resources efficiently, minimising waste and maximising utility. (Positive.)
3. *Efficiency in removing the intermediary.* There are no intermediaries involved in this transaction, eliminating administrative costs and potential inefficiencies associated with an intermediary's involvement. (Positive.)
4. *Efficiency in minimising market distortion and deadweight loss.* This scenario avoids market distortions because transactions are made privately and directly between private entities, eliminating the need for any taxes or subsidies which can potentially distort the market, deter trade, reduce gains from trade, and create deadweight losses. This scenario preserves market efficiency and maximises total surplus. (Positive.)
5. *Preference-compatibility for the taker.* Since you are both the spender and the beneficiary, there is perfect alignment between your spending and your needs and preferences, leading to maximum satisfaction and utility. (Positive.)

### 4.1.3 | Economic system attribution

This case resembles what occurs under a laissez-faire, free-market capitalist system, where individuals make and earn their own money and have full freedom to make their own economic decisions without interference. It reflects the principles of personal freedom and efficiency that characterise such a system, resulting in optimal allocation of resources and high levels of individual satisfaction.

## 4.2 | Case 2: Direct spending with restricted discretion

### 4.2.1 | Description

In this scenario, you spend your own money on yourself, but your spending choices are constrained by preset guidelines that require the money to be allocated only to specific predefined activities. While you have control over the spending, the lack of freedom to choose the exact use of your funds introduces limitations to your preference compatibility as a user.

### 4.2.2 | Criteria evaluation

1. *Preference-compatibility for the giver.* Since you are voluntarily allocating your own money to a need that you prefer, this scenario is preference-compatible for you as a giver. (Positive.)
2. *Efficient use of the transferred resource.* Since the money you are spending is yours and you are also the final taker, you are motivated to use it efficiently and avoid waste, ensuring optimal use of resources. (Positive.)
3. *Efficiency in removing the intermediary:* No intermediaries are involved in the transaction, which removes administrative costs and inefficiencies. (Positive.)

4. *Efficiency in minimising market distortion and deadweight loss*. Transactions occur directly, without the distortions introduced by taxes or subsidies, preserving market efficiency and maximising total surplus. (Positive.)
5. *Preference-compatibility for the taker*. Despite spending your own money on the cause you are interested in, the restrictions imposed on how you can use your own money in that cause makes you unable to choose how to use it, meaning that your preferences are not fully met, leading to suboptimal satisfaction and utility. (Negative.)

### 4.2.3 | Economic system attribution

This case resembles a scenario in an economic system where there are strict regulatory guidelines on how individuals can use their own money, such as in certain controlled economies or sectors with regulations. This case is very similar to laissez-faire capitalism but incorporates paternalistic controls over the type of consumption. In this scenario, you get to spend your own money on yourself, but you are restricted in your spending choices by certain preset guidelines or specific coupons. For instance, your income might be paid in the form of vouchers or coupons that can be used only for predetermined products, rather than cash that you can spend as you wish or on anything you would like to. In this case, individuals know the money is theirs, feel connected to it, and are motivated to use it wisely, but they lack the freedom to make consumption choices that closely match their true preferences. Other examples include 'government-mandated savings plans' and 'paternalistic-spending approaches with respect to one's own income'. This type of system maintains the core elements of a laissez-faire economy – such as personal financial responsibility and minimal intermediary involvement – but removes the freedom of choice in consumption. Thus, this case can be viewed as laissez-faire capitalism with the addition of consumption-specific coupons, where personal financial incentives are preserved, but individual choice in spending is constrained. While the spending is still direct and efficient in many ways, the lack of personal discretion mirrors situations where individual freedom of choice is curtailed.

## 4.3 | Case 3: Giving with recipient discretion

### 4.3.1 | Description

In this scenario, you voluntarily give your own money to someone else, and the recipient has the freedom to choose how to spend it. This set-up allows the giver to contribute willingly, while the recipient retains autonomy over the spending decisions, leading to a full preference compatibility with a mix of efficient and inefficient outcomes, as explained below.

### 4.3.2 | Criteria evaluation

1. *Preference-compatibility for the giver*. You are giving money voluntarily, aligning with your preferences and moral choices, ensuring a high degree of satisfaction. (Positive.)
2. *Efficient use of the transferred resource*. Since the giver and the recipient are two different entities, the efficient use of the resource may be jeopardised, regardless of which party engages in the spending process. The giver has no stake in the efficient use of the resources,[7] as they are not the final beneficiary, and also the recipient may not fully appreciate the effort required to earn the money, potentially leading to less efficient use than in the first two cases. In the first two cases, the fact that the giver, spender, and recipient were the same person made those scenarios ideal for utilisation efficiency. However, in Case 3 utilisation efficiency tends to be lower for the reasons explained previously. (Negative.)

3. *Efficiency in removing the intermediary.* The transaction occurs directly between the giver and the recipient, avoiding the inefficiencies and costs associated with intermediaries. (Positive.)
4. *Efficiency in minimising market distortion and deadweight loss.* Direct transactions preserve market efficiency by avoiding distortions caused by taxes or subsidies, maximising the total surplus. (Positive.)
5. *Preference-compatibility for the taker.* The recipient has the freedom to spend the money according to their preferences and needs, leading to high satisfaction and utility for the taker. (Positive)

### 4.3.3 | Economic system attribution

This case resembles charitable donations or financial gifts where the giver provides funds, but the recipient has full discretion over how to use them. Such scenarios are common in philanthropic activities, personal gifting, and certain educational scholarships, reflecting a spending mechanism that values personal choice and direct assistance. However, the lack of stake in utilisation and lack of appreciation of effort to earn the money can sometimes lead to inefficiencies in how the funds are used, which is a notable aspect of such spending mechanisms.

## 4.4 | Case 4: Giving with restricted recipient discretion

### 4.4.1 | Description

In this scenario, you spend your own money voluntarily on someone else, but the recipient does not have the freedom to choose how to spend it. The giver decides the specific use of the funds, which can lead to misalignment with the recipient's preferences.

### 4.4.2 | Criteria evaluation

1. *Preference-compatibility for the giver.* You are giving money voluntarily, aligning with your preferences and moral choices, ensuring a high degree of preference compatibility for you as the giver. (Positive.)
2. *Efficient Use of the transferred resource.* The giver has no stake in the efficient use of the resources, and the recipient may not fully appreciate the effort required to earn the money, leading to less efficient utilisation. (Negative.)
3. *Efficiency in removing the intermediary.* The transaction occurs directly between the giver and the recipient, avoiding the inefficiencies and costs associated with intermediaries. (Positive.)
4. *Efficiency in minimising market distortion and deadweight loss.* Direct transactions preserve market efficiency by avoiding distortions caused by taxes or subsidies, maximising the total surplus. (Positive.)

5. *Preference-compatibility for the taker.* The recipient's inability to choose how and on what items to spend the money may result in a mismatch with their preferences and needs, leading to preference incompatibility for the user. (Negative.)

### 4.4.3 | Economic system attribution

This case resembles scenarios where donations or financial assistance are provided with specific conditions on how the money must be used, such as scholarships with designated fields of study or grants with strict spending guidelines. Such situations are common in some philanthropic activities. Another example of such a spending mechanism is the use of coupons issued by private entities. The lack of freedom of choice and the lack of efficiency in utilisation can lead to reduced alignment with the recipient's needs and lowered levels of efficiency, reflecting an economic system that imposes conditions on financial support.

Cases 5–8 are symmetrical to Cases 3–4 and only view the situation from a different perspective. Similarly, Cases 9–10 are symmetrical to Cases 11–12 and only view the situation from a different person's perspective. As such, these symmetrical cases are not elaborated further here. The inclusion of these symmetrical cases in the model is necessary, for three reasons. First, it provides a comprehensive view of the full picture and the symmetries involved. Second, it allows us to fully observe the gradual shrinkage of the economic pie, freedom pie, and utility pie as the form of an economic system transition away from laissez-faire capitalism to pure communism. Third and most importantly, including these symmetrical cases enables the creation and discussion of Cases 3–4 in the first column and Cases 11–12 in the third column of Figure 4. Removing these symmetrical cases and their corresponding columns and rows would eliminate the second column, thereby removing Cases 3–4 and Cases 11–12, which are critical aspects of the extended version of Friedman's spending typology matrix. Therefore, all cases are preserved in the visual for the completeness of the analysis and depiction.

## 4.5 | Case 11: Indirect allocation through a third party with recipient discretion

### 4.5.1 | Description

In this scenario, a third-party intermediary, such as a government, allocates someone else's money to a recipient who has the freedom to choose how to spend it. This situation introduces inefficiencies and preference incompatibility for the original contributor, but allows the recipient to align the spending with their preferences.

### 4.5.2 | Criteria evaluation

1. *Preference compatibility for the giver.* The original contributor is often legally required to contribute, as in the case of taxes, and lacks control over how the funds are used, resulting in a misalignment with their preferences and priorities. (Negative.)
2. *Efficient use of the transferred resource.* The original contributor does not have control over how the money is to be spent, the spender (i.e. the intermediary) may lack a stake in the efficient use of the resources, and the intermediary and the recipient may not fully appreciate the effort required to earn the money, most likely leading to suboptimal usage, waste, and inefficient utilisation. (Negative.)

3. *Efficiency in removing the intermediary.* The involvement of a third-party intermediary, such as a government, introduces administrative costs, leakages, corruptions, and other such potential inefficiencies. (Negative.)

4. *Efficiency in minimising market distortion and deadweight loss.* The presence of taxes, subsidies, and other forms of intermediary involvement and government intervention can distort the natural state of market operations, leading to deadweight losses and reduced overall market efficiency. (Negative.)

5. *Preference-compatibility for the taker.* The recipient has full discretion to spend the funds according to their needs and preferences, resulting in a high level of satisfaction and utility. (Positive.)

### 4.5.3 | Economic system attribution

This case resembles scenarios where government programmes allocate funds to individuals, who then have the autonomy to decide how to use them. Examples include government aid programmes, governmental grants, Universal Credit, or universal basic income (UBI) initiatives. While these systems ensure that recipients can address their personal needs and preferences, the lack of involvement and stake by the original contributors, coupled with potential market distortions and administrative inefficiencies, highlight the challenges in such spending mechanisms. This scenario reflects aspects of welfare-state economies with UBI initiatives, where redistributive policies aim to support individual preferences but may face inefficiencies and misalignments with the contributors' values.

While welfare-state economies such as social democracies (e.g. the Nordic countries) strive to improve preference compatibility for the givers (i.e. taxpayers) through democratic processes – where those politicians whose policies align better with public preferences would receive more votes to hold governmental offices – these efforts also involve the engagement of various additional institutions such as a well-functioning justice system, free press, transparent governance, and a system of checks and balances in order to prevent inefficiencies, corruption, and the 'leaky bucket' issue. However, most of these measures introduce additional inefficiencies in terms of Criterion 3 (intermediation, administrative costs, and monitoring costs) and Criterion 4 (taxes and deadweight loss), because larger government, increased intervention, and higher taxes become necessary, all of which can increasingly distort free-market equilibrium outcomes.

## 4.6 | Case 12: Indirect allocation through a third party with recipient restrictions

### 4.6.1 | Description

In this scenario, a third-party intermediary, such as a government, allocates someone else's money to a recipient who does not have the freedom to choose how to spend it. This set-up creates a dual disconnect,[8] resulting in all forms of inefficiencies and preference incompatibilities for both the original contributors (e.g. the taxpayers) and the recipients of the government spending.

### 4.6.2 | Criteria evaluation

1. *Preference-compatibility for the giver*. The original contributor is often compelled by law to contribute, as in the case of taxes, and does not have control over how the funds are utilised, leading to a mismatch with their preferences and priorities. (Negative.)
2. *Efficient use of the transferred resource*. The original contributor is disconnected from the funds after contributing, and the third-party spender lacks a stake in the efficient use of the resources, and the recipient may not fully appreciate the effort required to earn the money, all of which can result in inefficient usage. (Negative.)
3. *Efficiency in removing the intermediary*. The involvement of a third-party intermediary, such as the government, introduces administrative costs and potential inefficiencies. (Negative.)
4. *Efficiency in minimising market distortion and deadweight loss*. The presence of taxes, subsidies, and other forms of government intervention in the market can distort the market, leading to deadweight loss and reduced overall market efficiency. (Negative.)
5. *Preference-compatibility for the taker*. The recipient is unable to choose how to spend the money, resulting in a misalignment with their needs and preferences, leading to suboptimal satisfaction and utility. (Negative.)

### 4.6.3 | Economic system attribution

This case resembles scenarios in centrally planned economies (such as the communist system)[9] where the government allocates resources with strict conditions on their use. It also parallels coupon- or quota-based distribution mechanisms, where recipients have no discretion over their purchases. Instead, they must use their coupons to buy the specific products for which the coupons are issued. In fact, the coupons specify exactly what can be purchased or consumed. This means that individuals receiving these resources cannot choose how to best meet their personal needs or preferences. For example, in a quota-based system, a family might receive a fixed amount of food staples regardless of their dietary preferences or requirements. Similarly, in a coupon system, individuals might receive vouchers that can be used only for specific items or services, limiting their ability to make spending decisions that align better with their unique circumstances. These mechanisms are often found in command economies or heavily regulated economies, where the government seeks to control and direct economic activity, often at the cost of individual freedom and efficiency. Such systems are characterised by rigid, bureaucratic structures where both efficiency and preference compatibility are compromised due to the dual disconnect between contributors and recipients, as well as the inefficiencies introduced by intermediaries and market distortions.

In conclusion, the examination of these 12 cases provides a detailed analysis of how different economic systems and various spending scenarios align with the five key outcome criteria studied in this article. Each scenario highlights the unique advantages and limitations inherent in various economic systems and spending mechanisms. To synthesise these findings, Figure 5 presents a comprehensive table summarising the advantages of each economic system and spending mechanism. This visual assigns numerical values between 0 and 5 to each scenario, with each point representing a positive outcome in one of the five criteria. This figure provides a straightforward way to compare the number of desirable criteria achieved under each scenario, offering valuable insights into the overall efficiency and preference compatibility of different spending approaches.

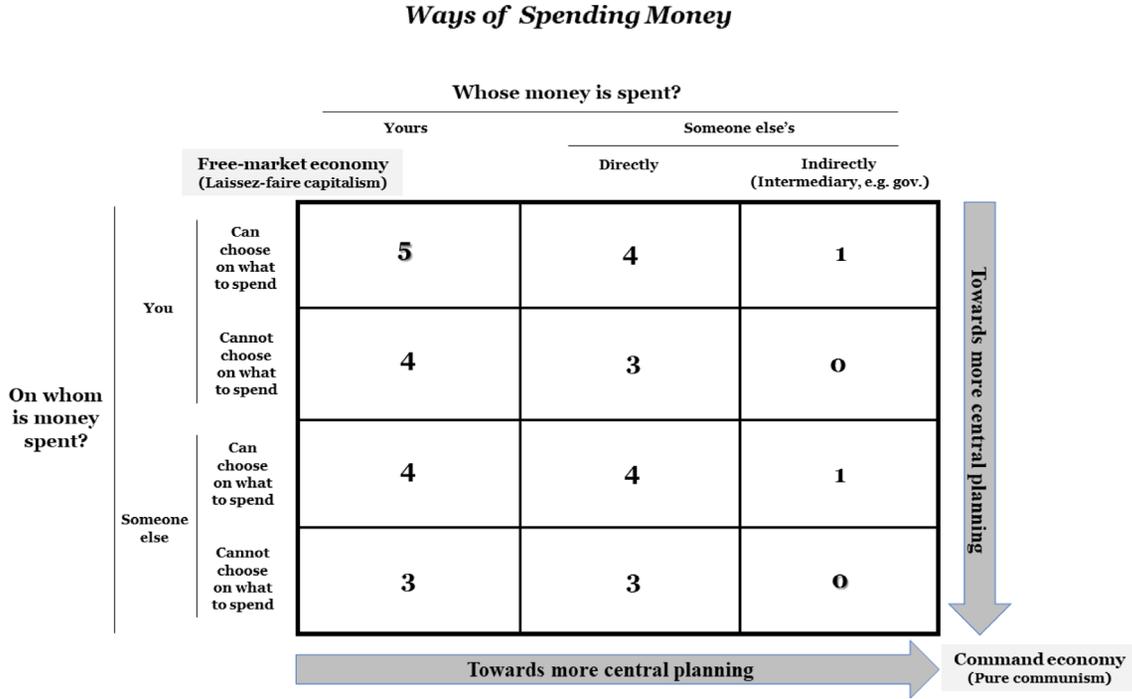

**FIGURE 5**   An extended version of Friedman's spending typology with scores associated with the overall advantages of each economic system and spending scenario.

Figure 5 provides an extended version of Friedman's spending typology by categorising 12 distinct cases of spending scenarios. Each case is evaluated based on five key outcome criteria: preference compatibility for the giver, efficient use of resources, removal of intermediaries, market efficiency, and preference compatibility for the taker. The figure allows for an easy assessment of how well each spending scenario performs in terms of efficiency and preference alignment.

As explained earlier, the upper-left corner of this figure (Case 1) represents laissez-faire capitalism, while the lower-right corner (Case 12) represents pure communism. The cases situated between these two extremes reflect intermediate forms of economic organisation, each incorporating elements of both free-market and command economies.[10] These intermediate cases demonstrate varying degrees of preference compatibility and efficiency, depending on their proximity to either end of the spectrum. The scoring system, as detailed above, allows the table to clearly illustrate the gradual deterioration of outcome criteria as an economic system's morphology transitions from laissez-faire capitalism to a fully command economy. This pattern is visually represented by the movement from the upper-left to the lower-right corner, highlighting how economic dynamics and efficiencies are impacted by increased regulation and reduced individual liberty and market freedom.

In essence, this extended version of Friedman's classification of ways of spending offers an insightful analysis of different economic systems based on their efficiency and preference compatibility. Efficiency, reflected in the size of the economic pie, indicates how well resources are used to maximise output and minimise waste. Preference compatibility, represented by the size of the freedom pie, illustrates the degree of freedom individuals have in making choices that align with their personal preferences. By scoring each scenario against these criteria, the figure provides a visualisation of how different economic systems balance efficiency and freedom of choice. This approach allows for a nuanced understanding of the trade-offs and dynamics that define various economic systems, highlighting the impact of regulation and market freedom on overall economic and individual well-being. As depicted in the figure, laissez-faire capitalism tends to deliver the highest degree of efficiency and preference compatibility, scoring five positive points, while pure communism tends to achieve the lowest levels of these outcomes, scoring zero positive points.

Indeed, the sense of dissatisfaction that people often experience under central planning stems from the multiple forms of inefficiencies, misalignments, and unmet preferences that arise as an economic system shifts away from a decentralised structure towards more centralised control. While it might be tempting to believe that manipulations and manual adjustments by a human-based, centralised institution or a social planner, such as the government, can improve the natural state of affairs, this idea has consistently proven flawed for complex, large-scale natural systems like the economy and nature. Such systems possess inherent self-regulating mechanisms that delicately balance and guide internal forces in very complex ways that are difficult to even fully explain, let alone optimise centrally. Just as humankind has accepted the self-regulating nature of ecosystems, there is fortunately a growing recognition that economies, likewise, operate best when they follow their own natural courses with minimal government interference. Although there are instances where central planning can be beneficial, such as in addressing negative externalities and market power, these cases are relatively rare compared with the vast scale of economic activity within a country. Therefore, the scope of government intervention should always be kept minimal to preserve the natural self-correcting processes that foster economic efficiency, freedom of choice, and overall satisfaction in society in the most effective manner.

**ECONOMIC AFFAIRS** $-$WILEY $-$ 

As Friedman (1962) famously stated, "Underlying most arguments against the free market is a lack of belief in freedom itself" (Friedman, 1962, p. 15). This freedom of choice, both for the giver in deciding how to allocate their resources and for the recipient in choosing what products to buy, is crucial for maximising the overall utility pie in the economy. This type of freedom is not only instrumental, enhancing economic efficiency and individual satisfaction, but also intrinsic, as Amartya Sen (1999) has emphasised that freedom is in fact a core value in its own right. Moving from the upper-left corner of our figure towards the lower-right corner, as Hayek (1944) described, represents "a shift from freedom to serfdom". Hayek (1944) warned that "the more the state 'plans' the more difficult planning becomes for the individual" (Hayek, 1944, p. 57), illustrating the dangers of excessive government intervention and centralised control. This transition captures the essence of two ends of a spectrum: at one end, freedom, individual responsibility, self-regulation, and market organisation, and at the other, coercion, state responsibility, government intervention, and central planning. As both Friedman and Hayek articulated, preserving economic freedom is essential for maintaining an efficient, preference-compatible system that maximises both economic and utility pies.

While Friedman's typology offers a valuable framework for analysing spending efficiency and preference compatibility, certain exceptions challenge the assumption that the spender always knows best. In cases involving individuals with cognitive impairments, addiction, self-control issues, or developmental immaturity (such as childhood), external decision-makers may be better equipped to allocate resources in ways that enhance long-term well-being. This perspective aligns somewhat with the concept of libertarian paternalism introduced by Thaler and Sunstein in their book *Nudge* (2008), which advocates interventions that guide individuals towards better choices without eliminating autonomy. Furthermore, preference compatibility and spending efficiency are highly dependent on the relationship between the giver and the recipient. When givers have close, personal knowledge of recipients – as in family or charitable settings – preference alignment is more likely. In contrast, public spending by distant intermediaries often suffers from preference misalignments and informational asymmetries and inefficiencies, as emphasised by Buchanan's public choice theory. Finally, while private spending is typically more efficient, certain goods – such as national defence – are inherently public in nature and cannot be efficiently supplied through market mechanisms due to their non-excludability and non-rivalry. Samuelson (1954) provided the foundational rationale for public provision of such goods, and Buchanan and Tullock (1962), along with Buchanan (1965), extended this logic to club goods, which may require a blend of state oversight and market mechanisms.

While the government plays an essential role in the economy in certain areas, it is important to recognise and always remind ourselves of the fact that, government "even in its best state, is but a necessary evil", as Paine (1776) said. Government intervention, though sometimes indispensable, as noted earlier, has the potential to be the worst type of intermediation in the economy. It can become authoritarian, exercising undue control over resources and fostering corruption. Unlike private entities, the government lacks the natural incentives to be cost-effective and efficient and also can evade responsibility when it owns the factors of production.

In essence, although the government is indispensable for maintaining order, enforcing property rights, providing public goods, protecting common resources, establishing a legal framework that ensures ethical, fair, and transparent market interactions, maintaining competition, and playing some other necessary public roles,[11] its inherent flaws and potential for misuse necessitate a cautious approach to its role in economic and social systems.

# 5 | SUMMARY AND CONCLUSION

This article has explored and extended Milton Friedman's original spending typology by conducting a comprehensive analysis of 12 distinct spending scenarios, each evaluated against five key outcome criteria. The extended typology enables comparison of the efficiency and preference compatibility of various spending mechanisms.

The analysis indicates that laissez-faire capitalism consistently achieves the highest overall score in both efficiency and preference compatibility. This system benefits from minimal government intervention, allowing individuals to make autonomous decisions that align closely with their preferences and optimise resource utilisation. In contrast, pure communism unsurprisingly scores the lowest due to extensive government control, which often results in various forms of inefficiencies and misalignments with individual preferences.

The examination of intermediate cases between these two extremes underscores the complexities and varied outcomes associated with mixed economic systems. While certain scenarios, such as charitable donations with recipient discretion, highlight the benefits of decentralised decision-making, others, such as government-mandated spending allocations, expose significant inefficiencies and dissatisfaction. The analysis also highlights the gradual deterioration of advantages in spending mechanisms as we move away from laissez-faire capitalism toward pure communism, emphasising the increasing inefficiencies and preference misalignments associated with greater government intervention and control.

These findings align with the broader economic arguments made by Friedman and Hayek, emphasising the critical importance of economic freedom and minimal government intervention. As Friedman once noted, "When government—in pursuit of good intentions—tries to rearrange the economy, legislate morality, or help special interests, the cost comes in inefficiency, lack of motivation, and loss of freedom" (Friedman & Friedman, 1980, p. 148). The extended model clearly illustrates how government interventions lead to a diminished economic and utility pie, validating Friedman's warning about the unintended consequences of government overreach.

In conclusion, while government intervention and non-market spending mechanisms are sometimes necessary to address specific market failures, government's role should be carefully limited to avoid the pitfalls of inefficiency and preference misalignment. The delicate balance between necessary government involvement and the preservation of economic freedom is crucial for fostering a prosperous and just society. This article suggests that the most efficient and the most preference-compatible economic systems are those that empower individuals with the personal freedom to make their own choices.

## ENDNOTES

[1] The metaphor of the 'leaky bucket', owed to Arthur Okun (2015) describes the inefficiencies and losses that occur during the transfer of resources through intermediaries, such as bureaucratic processes and administrative costs. Okun uses this metaphor to illustrate how, in the process of redistributing income, some of the resources are inevitably lost due to administrative expenses, leakage, and other such inefficiencies.

[2] It is also important to note that, while intermediaries usually introduce inefficiencies, they may sometimes add value by reducing transaction costs or utilising specialised knowledge under certain circumstances (e.g. you can see Coase (1937) and GiveWell (https://www.givewell.org/), which is a charity organisation, serving as an intermediary to evaluate various forms of charities and recommend cost-effective forms of charities). The concept of intermediation inefficiency as it relates to administrative expenses, leakage, and corruption is directly connected to the third outcome criterion, which highlights the efficiency gained by removing

intermediaries and minimising the 'leaky bucket' effect to ensure that more resources reach their intended recipients without unnecessary waste or corruption.

[3] It is also important to acknowledge that deadweight loss can arise in various forms, even within private transactions, such as in the case of monopolies or other forms of market failures, which are usually rare in the large scale of economic activity within free markets. Although deadweight loss is not always eliminated fully in private transactions, it tends to be far less than occurs with forced or centralised allocations.

[4] The first and last criteria here can also be understood as two notions of efficiency in matching preferences with target goods and services. In matching theory, a sub-branch of game theory, efficiency pertains to the extent to which a matching or resource allocation maximises overall social welfare or utility. This concept of efficiency focuses on achieving the best possible outcomes within the constraints and available resources. In matching problems – such as pairing students with schools, workers with jobs, donations with needs, and citizens with the public goods they need – efficiency means creating optimal matches that best align individual preferences with available opportunities. By maximising efficiency, matching theory strives to ensure that allocations yield the highest overall satisfaction or social welfare among participants, thus enhancing efficiency and satisfaction (or possibly some other objectives such as stability and fairness) in the matching process.

[5] This point is crucially important because many philosophers, along with political and legal scholars, believe that the central issue in politics is the preservation and enhancement of human freedom. Consequently, autonomy in spending is essential for maintaining personal liberty and satisfaction.

[6] In judging which way of spending is desirable or undesirable, this article relies on equally weighted outcome criteria. However, one may opt for a more sophisticated assessment of relative importance and weights of the five criteria if there is a strong justification for attributing unequal weightings to these five factors.

[7] The phrase 'lack of stake' refers to a situation where an individual does not have a personal or significant interest or investment in the consumption or use of a particular product, service, or resource. It implies a lack of ownership, responsibility, or direct consequences associated with the consumption process. When someone lacks a stake in consumption, they may exhibit indifference or minimal concern about the quality, quantity, or impact of their consumption choices, which can lead to a decreased sense of responsibility or accountability for the outcomes or effects of their consumption behaviour. For example, if someone is using a communal facility or a shared resource where they do not have a direct financial or personal ownership stake, they might be less motivated to conserve or maintain the resource properly, resulting in careless usage or a disregard for the long-term sustainability and quality of the resource.

[8] The concept of the dual disconnect refers to two distinct layers of separation that contribute to inefficiencies and misalignments in preference compatibility in a spending scenario. Specifically, it involves: (*a*) Disconnect between the contributor and the intermediary. The original contributor (the person or entity providing the funds) has no control or influence over how the money is spent once it is handed over to a third-party intermediary, such as a government or charity organisation. This means the contributor's preferences and intentions are not taken into account in the actual spending process. (*b*) Disconnect between the intermediary and the recipient. The recipient of the funds does not have the freedom to choose how to use the money because the third-party intermediary either provides only one or a limited range of public goods or services, leaving no room for choice or opting out, or it imposes specific conditions or guidelines on fund usage. As a result, the recipient's needs and preferences are not fully met, leading to reduced satisfaction and utility. Together, these two disconnects create a situation where neither the preferences of the original contributor nor the needs of the recipient are fully addressed, resulting in significant inefficiencies, considerable preference misalignments, and suboptimal outcomes.

[9] In this article, the definitions of capitalism, Scandinavian socialism, and communism are in line with those introduced in Zeytoon-Nejad (forthcoming), which introduces a novel quantitative approach to classifying economic systems. That study develops institutional similarity indices – Capitalism Similarity Index (CapSI), Socialism Similarity Index (SocSI), and Communism Similarity Index (ComSI) – to objectively position countries along the economic-system continuum based on hard data rather than subjective categorisations.

[10] The extended spending typology in this article does not suggest a strict binary axis between laissez-faire capitalism and communism but rather illustrates general tendencies in spending efficiency and preference compatibility across different economic systems. It is acknowledged that real-world economies are inherently

hybrid, combining elements from multiple spending categories. Even market-oriented systems include some public spending, and centrally planned economies allow some private expenditure. The typology presented here reflects only the predominant economic philosophy and institutional logic behind each spending category.

[11] Indeed, these roles create the foundational conditions for economic stability, growth, and individual prosperity. While this article highlights the inefficiencies and preference misalignments that arise when governments act as intermediaries in economic transactions, it does not seek to understate the mentioned essential roles that governments must play in any well-functioning economy. However, beyond such necessary functions, excessive governmental involvement – particularly in areas where private decision-making and market mechanisms can operate efficiently on their own – often introduces bureaucratic inefficiencies, misaligned incentives, and unnecessary administrative costs.

**How to cite this article:** Zeytoon-Nejad, A. (2025). Milton Friedman's spending matrix revisited: 'Spending efficiency' and 'preference compatibility' across different economic systems. *Economic Affairs*, 1–24. https://doi.org/10.1111/ecaf.12700